# Electrical Properties of Sculpturenes

Laith Algharagholy[1], Thomas Pope[2], Steven W.D. Bailey[2] and Colin J. Lambert[2]*

[1]Computer Science and Mathematics College, University of Al-Qadisiyah, Diwaniyah, Iraq

[2]Department of Physics, Lancaster University, Lancaster LA1 4YB, UK

*Address correspondence to c.lambert@lancaster.ac.uk

**We investigate the electronic properties of *sculpturenes,* formed by sculpting selected shapes from bilayer graphene, boron-nitride or graphene-boron-nitride hetero-bilayers and allowing the shapes to spontaneously reconstruct. The simplest sculpturenes are periodic nanotubes, containing lines of non-hexagonal rings. More complex sculpturenes formed from shapes with non-trivial topologies, connectivities and materials combinations may also be constructed. Results are presented for the reconstructed geometries, electronic densities of states and current-voltage relations of these new structures.**

Key words: nanostructures, self-assembly, directed assembly, graphene, carbon nanotubes, boron nitride, nanoribbons

## 1. Introduction

Sculpturenes are novel nanometre-scale objects, obtained by cutting selected shapes from layered materials and allowing the shapes to reconstruct [1, 2].The simplest examples of sculpturenes are formed by cutting straight nanoribbons from bilayer graphene and allowing the edges to reconstruct to maximise sp2 bonding. If the width of the nanoribbon is sufficiently small (i.e. of order 3nm or less) then the whole ribbon can reconstruct to form a carbon nanotube, with a pre-defined location and chirality. More complex all-carbon structures with unique topologies are also possible, as are T-shapes, crosses and other multiply-connected structures. By sculpting hetero-bilayers such as graphene on monolayer boron nitride and allowing them to reconstruct, new hetero nanotubes can be made, as well as a variety of more complex geometries formed from two or more materials.

In practice, a variety of techniques are available for cutting sculpturenes [3]. All of these require specific experimental conditions and deliver cuts with different levels of accuracy. The atomic-scale dynamics of these methodologies are largely unknown and therefore in what follows, we circumvent this issue by starting from bi-layers with pre-cut edges and then using density functional theory (DFT) to allow them to reconstruct. Methods for achieve cut graphene include lithographic [4,5,6], chemical [7,8,9,10] and sonochemical [11,12] techniques. In particular, STM lithography [13] can be used to cut GNRs with widths as small as 2.5nm, with a specified chirality, a specified location and with their ends contacted to (graphene) electrodes and aberration-corrected TEM can be used to cut holes in graphene and other materials [14,15]

The aim of the present paper is to investigate the electron transport properties of a selection of these novel structures. Initially we shall examine the electronic properties of sculpturenes formed by cutting straight bilayer nanoribbons and allowing them to relax. Depending on the direction of cutting, the resulting sculpturenes can be either perfect nanotubes, or nanotubes with lines of non-hexagonal rings. Electronic properties of more complex sculpturenes are also presented, including their densities of states and electron transport properties.

## 2. Electronic and geometric properties of all-carbon sculpturenes formed from straight nanoribbons.

In what follows, all structures are relaxed using the SIESTA implementation of DFT [4] using the Ceperley-Alder (CA) exchange correlation functional, with norm-conserving pseudopotentials and double zeta polarized (DZP) basis sets of pseudo atomic orbitals.

Figure 1 shows examples of pre-cut, straight, bilayer graphene nanoribbons (BiGNRs), which are infinitely periodic in the horizontal direction and of finite width W in the vertical direction, along with the corresponding nanotubes, which form after allowing them to reconstruct spontaneously. The initial BiGNRs are AB-stacked bilayer zigzag graphene nanoribbons (BiZGNRs), which after relaxation form perfect armchair CNTs with no defects.

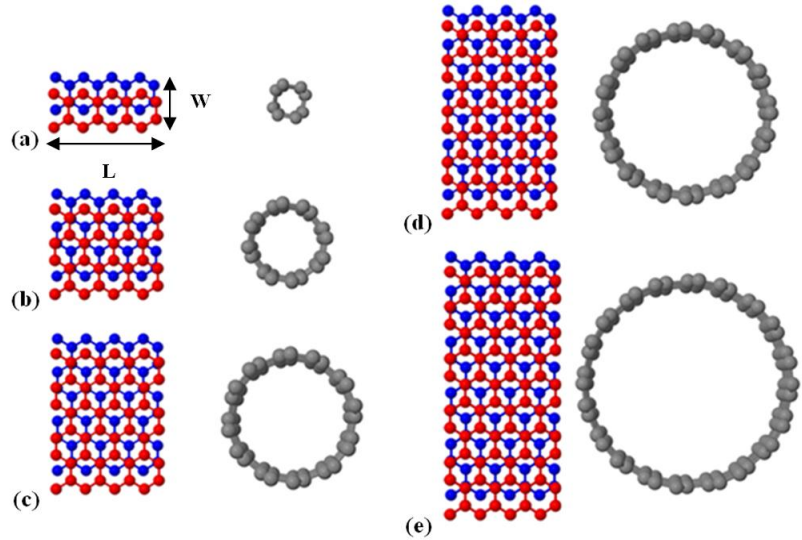

*Figure 1 Armchair CNTs obtained by relaxing AB-stacked BiZGNRs with different widths W. The ribbons are infinitely periodic in the horizontal direction and to allow the possibility of surface reconstruction, supercells of length L =3.5×3$^{1/2}$ a$_{c-c}$ are employed, where a$_{c-c}$ = 1.44 Å is the carbon-carbon bond length. (a) W=3 a$_{c-c}$ relaxes to a (2,2) armchair CNT. (b) W =6 a$_{c-c}$ relaxes to a (4,4) armchair CNT. (c) W =9 a$_{c-c}$ relaxes to a (6,6) armchair CNT. (d) W =12 a$_{c-c}$ relaxes to a (8,8) armchair CNT. (e) W =15 a$_{c-c}$ relaxes to a (10,10) armchair CNT.*

As further examples, figure 2 shows three sculpted AB-stacked bilayer armchair graphene nanoribbons (BiAGNRs), which after reconstruction form perfect zigzag CNTs.

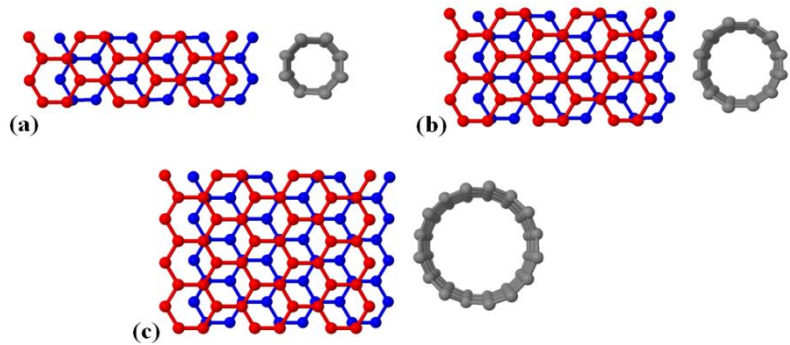

*Figure 2 Zigzag CNTs obtained by relaxing AB-stacked BiAGNRs with different widths W. The ribbons are infinitely periodic in the horizontal direction and to allow the possibility of surface reconstruction, supercells of length L =8a$_{c-c}$ are employed, where a$_{c-c}$ = 1.44 Å is the carbon-carbon bond length. (a) W=1.5×3$^{1/2}$a$_{c-c}$ relaxes to a (4,0) zigzag CNT. (b) W =2.5×3$^{1/2}$a$_{c-c}$ relaxes to a (6,0) zigzag CNT. (c) W =3.5×3$^{1/2}$a$_{c-c}$ relaxes to a (8,0) zigzag CNT.*

The above examples are only a subset of the CNTs that can be obtained by relaxing armchair or zigzag BiGNRs, because the resulting CNTs depend not only on the orientation of the BiGNR, but also on the combination of edges which coalesce during reconstruction. To illustrate how different combinations of edge terminations can affect the resulting CNTs, consider first the zigzag and armchair edge terminations of GNRs, shown in figure 3. Figures 3 (a-f) show six possible upper and lower edge combinations of monolayer zigzag-terminated GNRs (ZGNRs), formed from edge terminations labelled T$_1$, T$_2$, T$_3$ and T$_4$. Figures 3 (g-h) show two possible edge combinations of armchair-terminated GNRs (AGNRs) labelled T'$_1$ and T'$_2$.

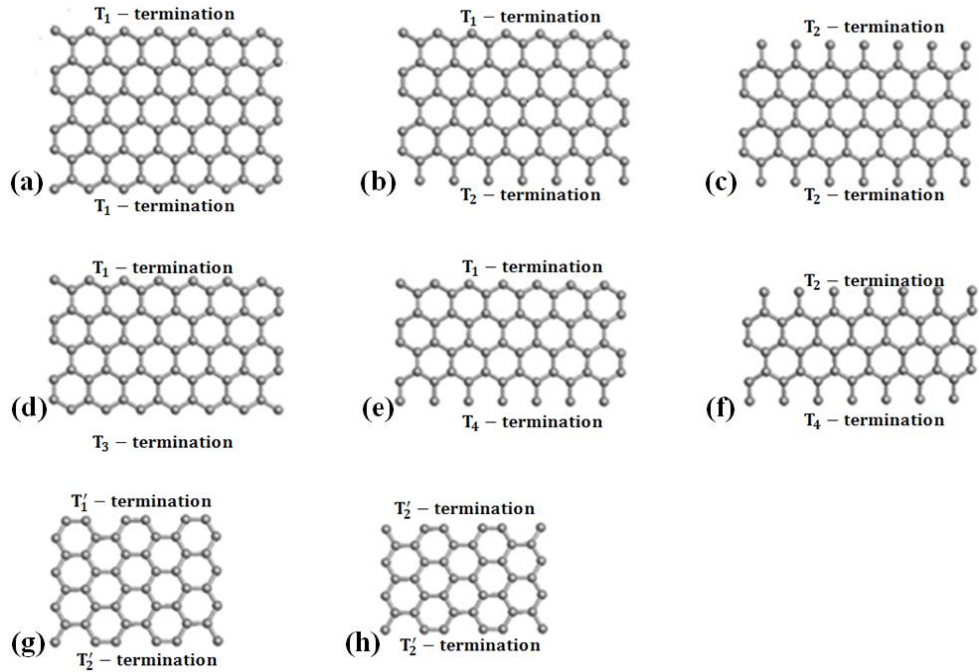

*Figure 3 Figs (a-f) Supercells of monolayer ZGNRs (periodic in horizontal direction) with various widths W and identical lengths. Six ZGNR terminations are shown, (a) T$_1$T$_1$-termination (84 atoms) with W=8 a$_{c-c}$. (b) T$_1$T$_2$-termination (77 atoms) with*

*W=7.5 $a_{c-c}$. (c) $T_2T_2$-termination (70 atoms) with W=7 $a_{c-c}$ (d) $T_1T_3$-termination (70 atoms) with W=6.5 $a_{c-c}$. (e) $T_1T_4$-termination (63 atoms) with W=6 $a_{c-c}$. (f) $T_2T_4$-termination (56 atoms) with W=7 $a_{c-c}$. Figs (g-h) show supercells of AGNRs with width W and identical lengths. Two terminations are shown. (g) $T'_1 T'_2$-termination (48 atoms) with W=3.5× $3^{1/2}a_{c-c}$. (h) $T'_2 T'_2$-termination (42 atoms) with W=3×$3^{1/2}a_{c-c}$.*

When an AB-stacked bilayer is cut to form BiZGNRs, the bilayer nanoribbons possess two upper and two lower edges associated with each of the stacked monolayer ribbons. Each pair of upper (or lower) edges can be formed from a combination of the edges shown in figure 3. For BiZGNRs, examples of these combinations are shown in figure 4 (a-f), along with the resulting CNTs following reconstruction. Each BiZGNR possesses a pair of upper-edge and lower-edge terminations. For example in figure 4 (b), these are $T_1T_1$ and $T_2T_2$ respectively.

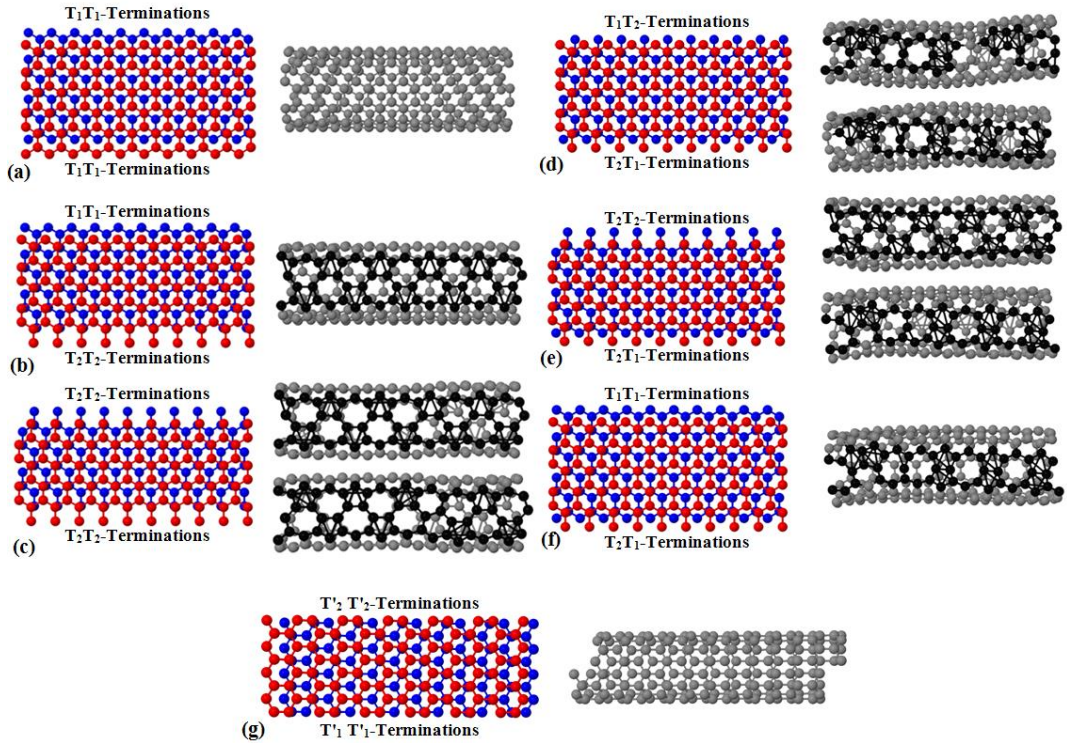

*Figure 4    Figs (a-g) show AB-stacked BiZGNRs with different edge terminations of the top ZGNR (red) and bottom ZGNR (blue). (a) Shows a supercell with $T_1T_1/ T_1T_1$ – terminations, comprising 240 carbon atoms. (b) $T_1T_1/ T_2T_2$ -terminations (220 carbon atoms). (c) $T_2T_2/ T_2T_2$ –terminations (200 carbon atoms). (d) $T_1T_2/ T_1T_2$-terminations*

*(220 carbon atoms). (e) $T_2T_2$/ $T_1T_2$-terminations (210 carbon atoms). (f) $T_1T_1$/ $T_1T_2$-terminations (230 carbon atoms). (g) A BiAGNR with $T'_1$ $T'_1$ / $T'_2$ $T'_2$-terminations (96 atoms) where the top AGNR is blue and the bottom AGNR is red. After reconstruction, the edges reconstruction is shown in black.  In general, we adopt the notation that a BiGNR with $T_iT_j$/ $T_kT_l$-terminations has a bottom (top) GNR with upper and lower edge terminations $T_i$ and $T_k$ ($T_j$ and $T_l$).*

The CNTs shown in figures 4 (a-g) are the relaxed structures resulting from each of their adjacent BiGNRs. To perform the geometry relaxation, it is necessary to define a supercell in the initial BiGNR. Since the reconstructed sculpturene may have a larger unit cell than the initial BiGNR, the supercell is chosen to be larger than the unit cell of the BiGNR and sufficiently large to accommodate the periodicity of the reconstructed sculpturene. Although this is not necessary for the sculpturenes of figures 1 and 2, it is clearly necessary for those of figure 4.  Figure 4 (g) shows that the $T'_1$ $T'_1$ / $T'_2$ $T'_2$-terminations relax to a perfect (8,0) zigzag CNT. Further terminations such as $T'_1$ $T'_2$ / $T'_1$ $T'_2$, $T'_1$ $T'_1$ / $T'_1$ $T'_2$ and $T'_2$ $T'_2$ / $T'_1$ $T'_2$  also form a perfect zigzag CNT . This demonstrates that the formation of zigzag CNTs from AGNRs is rather robust. Figure 4 (a) shows that $T_1T_1$/ $T_1T_1$ terminations of BiZGNRs relax to a perfect (6,6) armchair CNT. Figure 4 (b) shows that $T_1T_1$/ $T_2T_2$ terminations produce an armchair CNT with a line of pentagon-heptagon pairs. Figure 4 (c) shows that $T_2T_2$/ $T_2T_2$ terminations produce an armchair CNT with two lines of pentagons-heptagon pairs. Figure 4 (d) shows that $T_1T_2$/ $T_1T_2$ terminations lead to an armchair CNT with two lines of non-hexagonal rings, which contain octagons, horizontal pentagon-pairs and vertical pentagon-pairs. Figure 4 (e) shows that $T_2T_2$/ $T_1T_2$ terminations lead to two lines of non-hexagonal rings. In this case, the bottom line contains four octagons, one horizontal pentagon-pair and three vertical pentagon-pairs per supercell and the top line contains nine pentagons and heptagons. This tendency for polygons with more than 6 sides, to be attracted by polygons with less than 6 sides is a generic topologically-driven feature of networks of three-fold vertices [5, 6]. Figure 4 (f) shows that $T_1T_1$/ $T_1T_2$ terminations lead to an armchair CNT with one line of four octagons, one horizontal pentagon-pair and three vertical pentagon-pairs per supercell. Such lines of non-hexagonal rings are likely to possess novel spintronic and electronic properties [7-9].

We now investigate the electronic properties of the nanotube sculpturenes shown above, all of which are periodic in the horizontal direction, albeit in some cases with large unit cells (supercells). Using the DFT code SIESTA, results are obtained for the electronic density of states (DOS), the band structure and the number of open channels. Within a Landauer description of electron transport, since the CNTs investigated are periodic, the transmission coefficient T(E) for electrons of energy E travelling from left to right is identical to the number of open channels and therefore as an indicator of the maximum current I carried by the CNTs at a finite voltage V, we also present the quantity I(V) defined by:

$$I(V) = \frac{2e}{h} \int_{E_f - \frac{eV}{2}}^{E_f + \frac{eV}{2}} T(E) dE \qquad (1)$$

As initial examples, figure 5 shows the density of states, band structure and number of open channels for the three CNT sculpturenes in figures 4 (a), (c) and (g), which possess rather short unit cells in the horizontal direction. The sculpturenes shown in figures 5a and 5c are ideal nanotubes and possess the DOS of armchair and zigzag nanotubes respectively. In contrast, the CNT sculpturene shown in figure 5b contains a periodic line of 5/7 rings and the resulting DOS contains an energy gap near the Fermi energy of approximately 0.25 eV. This persistence of the energy gap is also found in CNTs containing periodic "zips" of impurities [10, 11]. In all cases, the DOS contains van Hove singularities associated with the periodic nature of these quasi-one-dimensional structures.

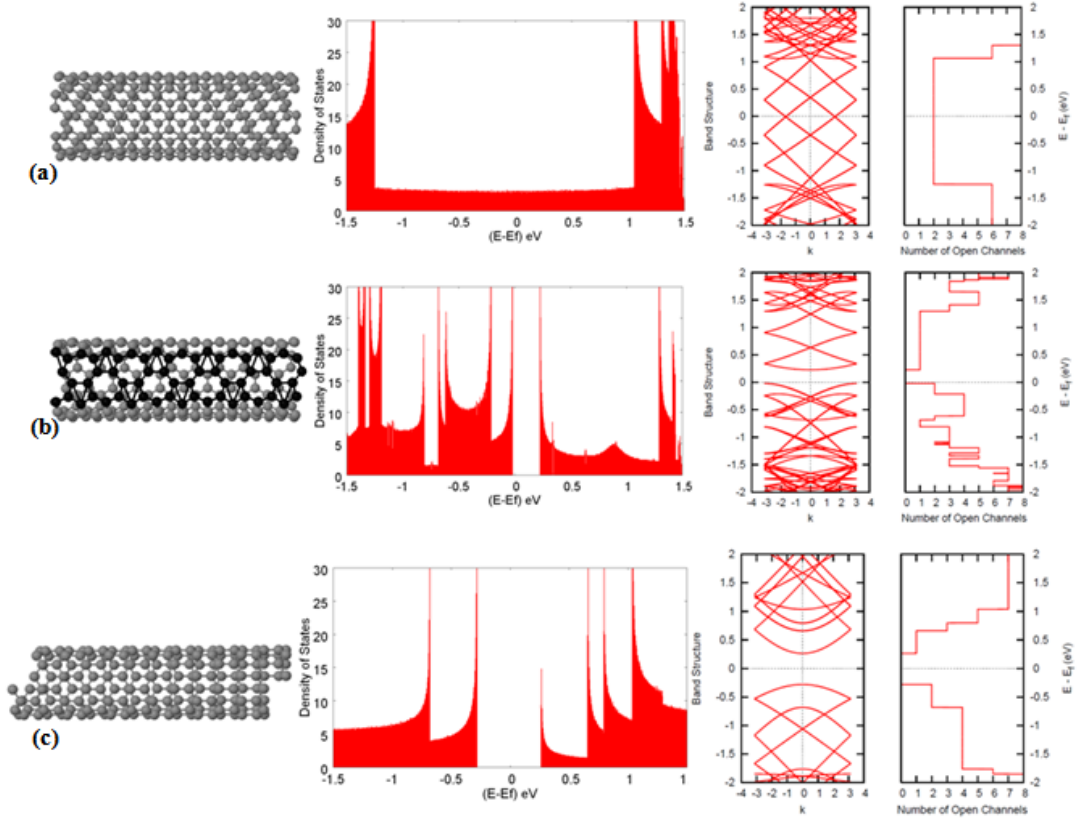

*Figure 5: Shows the density of states, band structure and number of open channels for ordered CNT sculpturenes. (a) A perfect (6,6) CNT sculpturene. (b) A CNT sculpturene with one line of 5/7pairs. (c) A perfect (8,0) CNT sculpturene. The density of states carried out with 3000 k-points.*

The band structures and the number of open channels (equal to the number of mini-bands) at energy E for each system are shown in the far right of the figures. As expected, for the perfect (6,6) nanotube, we find a large interval of energy with two open channels near the Fermi energy, whereas the zigzag (8,0) nanotube possesses a gap at the Fermi energy[12]. The zipped nanotube with a periodic line of 5/7 rings possesses a smaller gap of 0.25eV and exhibits a jump to either 1 or 2 open channels at the upper and lower band edges closest to the Fermi energy. By identifying the transmission coefficient T(E) with the number of open channels, the I-V curves obtained from equation (1) are shown in figure 6. It is clear from these that the maximum current through the 5/7-zipped nanotube lies between that of the perfect zig-zag and armchair CNTs.

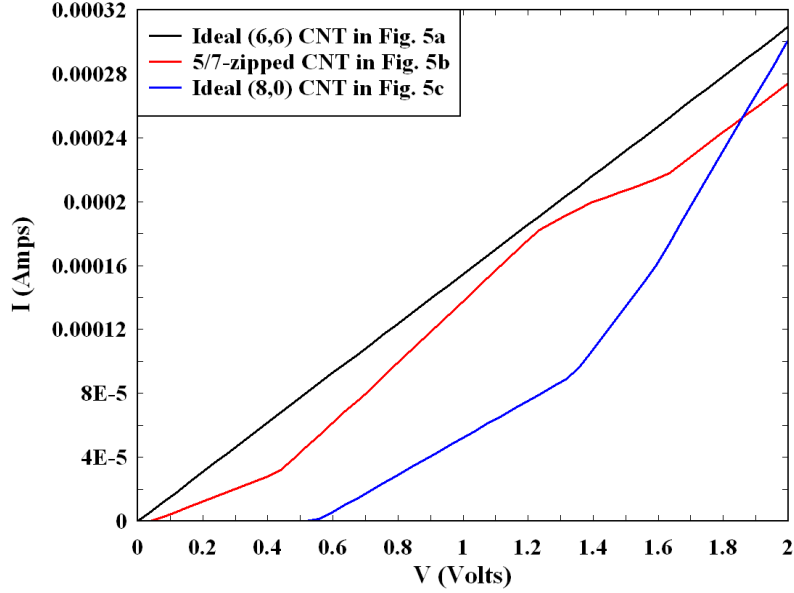

*Figure 6: Shows I-V curve of the ordered CNT sculpturenes in figures 5 (a, b and* c).

The DOS of examples of CNT sculpturenes with more complex unit cells are shown in figures 7. Figure 7a shows the CNT sculpturenes of figure 4c, one side of which possesses a line of with nine pentagon/heptagon pairs and a diametrically-opposite side containing a line of eight pentagon/heptagon pairs per supercell. Figure 7c shows the CNT sculpturene of figure 4e, one side of which possesses a line of nine vertical pentagon/heptagon pairs per supercell, while the opposite side contains four octagons, one horizontal pentagon-pair and three vertical pentagon-pairs per supercell. These show that in contrast with the energy gap shown in figure 5b there is almost no gap at Fermi energy and there is a corresponding increase in the electronic states around the Fermi energy.

For the sculpturene shown in figure 7b, the DOS shows many mini energy gaps around Fermi energy. This corresponds to the CNT sculpturenes of figure 4d. Similarly, figure 7d corresponds to a CNT sculpturene with one line of four octagons, one horizontal pentagon-pair and three vertical pentagon-pairs. The presence of quasi-disordered supercells in these sculpturenes leads to a significant increase in the DOS around the Fermi energy[13].

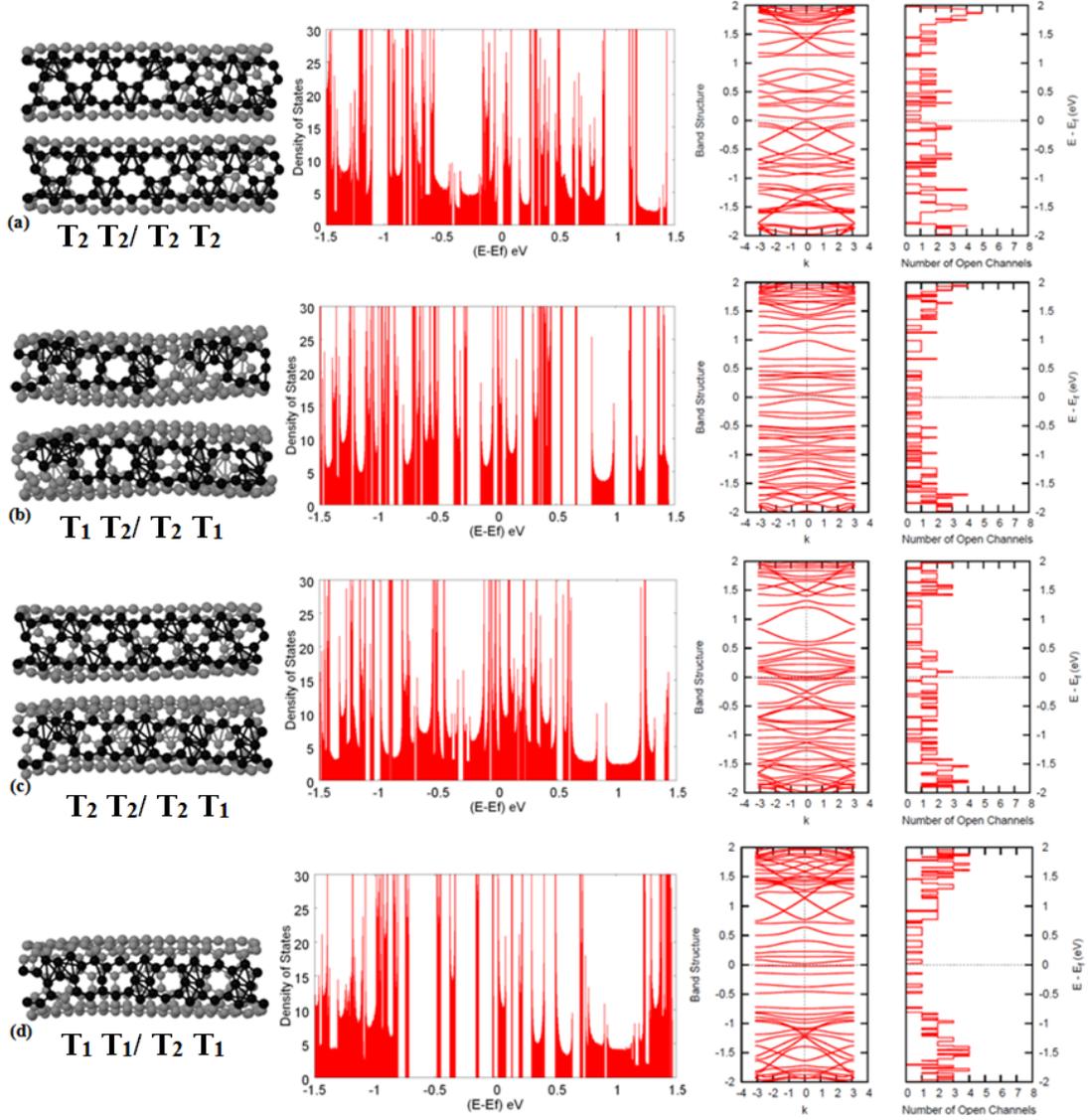

*Figure 7: Figures (a-c) Show the density of states, band structure and number of open channels for disordered CNT sculpturenes with two sides of non-hexagonal ring lines. (a) One side with nine vertical pentagon/heptagon pairs line and second side with eight vertical pentagon/heptagon pairs line. (b) The two sides with octagons, horizontal pentagon-pairs and vertical. (c) One side with nine vertical pentagon/heptagon pairs line and second side which contains line contains four octagons, one horizontal pentagon-pair and three vertical pentagon-pairs. (d) A CNT*

*sculpturene with one line of four octagons, one horizontal pentagon-pair and three vertical pentagon-pairs. The density of states carried out with 3000 k-points.*

From the number of open channels, it is clear that the conductance of sculpturenes with quasi-disordered unit cells is typically lower than that of conventional conducting nanotubes, because the number of open channels rarely exceeds 2. The I-V plots (figure 8) confirm that the current is lower than that of a comparable ideal (6,6) nanotube.

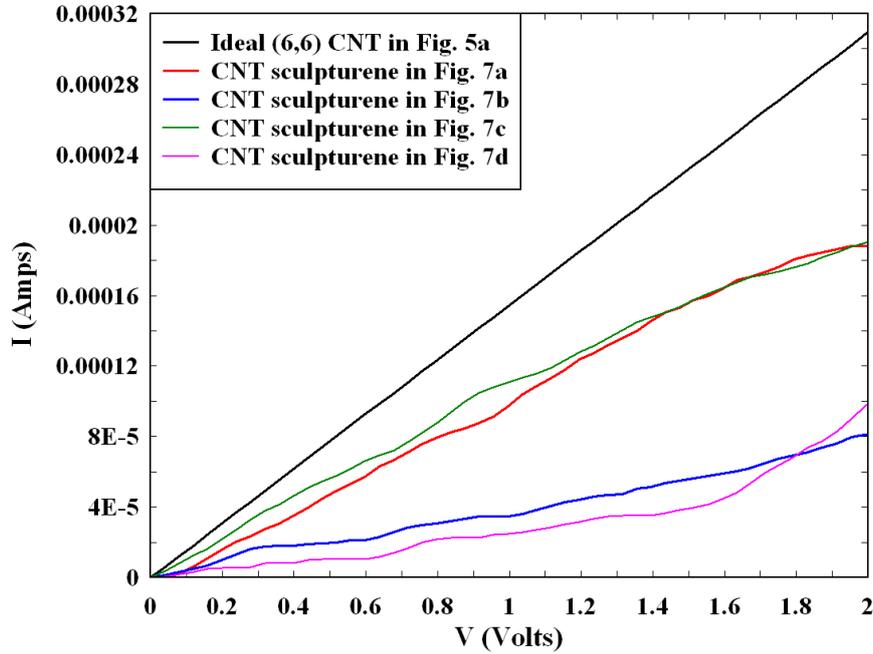

*Figure 8: Shows I-V curve of the ordered CNT sculpturenes in figures 7 (a, b, c and d).*

### 3. Electronic properties of BN sculpturenes formed from straight nanoribbons.

We now examine the electronic properties of BN sculpturenes. Conventional boron nitride nanotubes (BNNT) are quasi-one-dimensional nanostructures predicted in 1994 [14] and experimentally discovered in 1995. BN is an electrical insulator with a wide band-gap of approximately 5 eV [15-17]. In this section we examine the electronic properties of boron nitride nanotubes formed by relaxing bilayer BN nanoribbons. As examples, figure 9 shows three different combinations of edge terminations of zigzag BN nanoribbons.

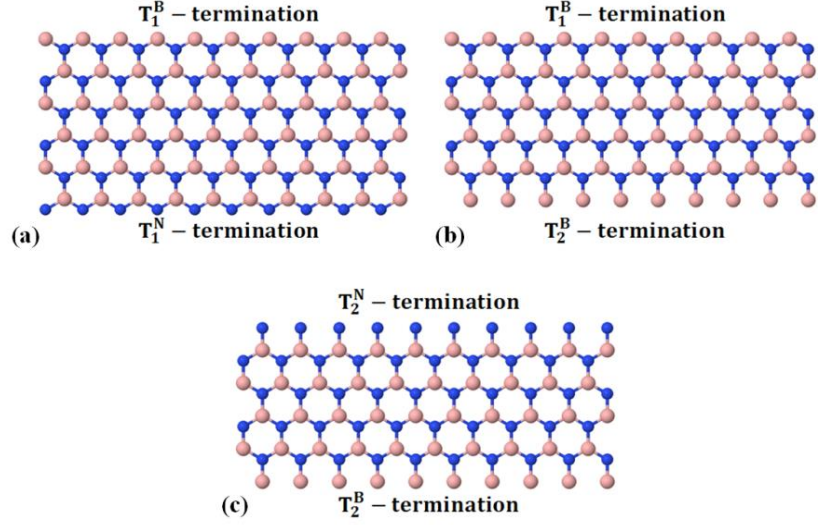

*Figure 9 Figs (a-c) supercells of monolayer zigzag boron-nitride nanoribbon BiZBNNR with various widths W and identical lengths. Three ZBNNR terminations are shown. (a) The top edge is terminated with boron $T_1^B$ and the bottom edge is terminated with nitrogen $T_1^N$ (120 nitrogen and boron atoms). (b) The top edge is terminated with boron $T_1^B$ and the bottom edge is terminated with boron $T_2^B$ (110 nitrogen and boron atoms). (c) The top edge is terminated with nitrogen $T_2^N$ and the bottom edge is terminated with nitrogen $T_2^B$ (100 nitrogen and boron atoms). For clarity, nitrogen atom is shown in blue and boron is shown pink.*

When an AB-stacked bilayer is cut to form bilayer zigzag boron-nitride nanoribbons, the bilayer nanoribbons possess two upper and lower edges associated with each of the stacked monolayer ribbons. Each pair of upper (or lower) edges can be formed from a combination of the edges shown in figure 9. For bilayer zigzag boron-nitride nanoribbons, examples of these combinations are shown in figure 10 (a-c), along with the resulting boron-nitride NTs following reconstruction. Each bilayer zigzag boron-nitride nanoribbon possesses a pair of upper-edge and lower-edge terminations, such as $T_1^N T_1^N$ and $T_1^B T_1^B$ in figure 10a.

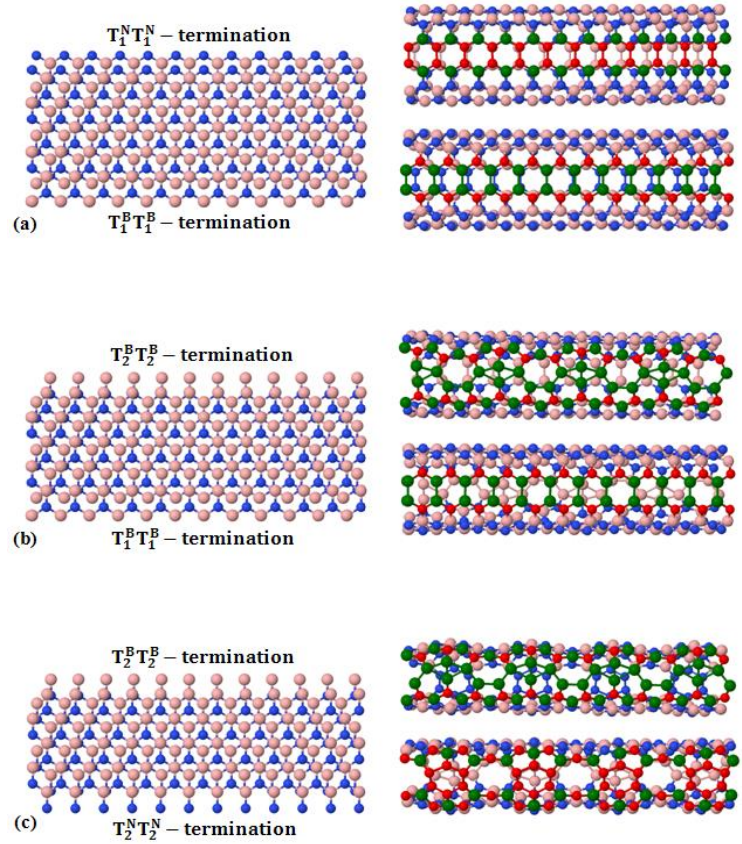

*Figure 10 The top left sub-figures (a-c) show a supercell of AB-stacked of bilayer zigzag boron-nitride nanoribbon BiZBNNRs (boron is shown in pink and nitrogen is shown in blue) with (a) $T_1^N T_1^N$ − termination (top edge) and $T_1^B T_1^B$ − termination (bottom edge), comprising 240 nitrogen and boron atoms while the top right sub-figure shows the relaxed boron-nitride NT sculpturene with two lines of hexagonal rings. In this case the top line contains hexagon rings zip, each hexagonal ring contains four nitrogen atoms (red) and two boron atoms (green) while the bottom line contains hexagon rings zip, each hexagonal ring contains four boron atoms (green) and two nitrogen atoms (red). (b) $T_2^B T_2^B$ − termination (top edge) and $T_1^B T_1^B$ − termination (bottom edge), comprising 220 nitrogen and boron atoms while the top right sub-figure shows the relaxed boron-nitride NT sculpturene with top line contains four-membered boron (three) rings, heptagons (four), horizontal pentagons pairs (three), pentagons (four) and horizontal hexagons pairs (three), the bottom line contains hexagon rings zip with four boron atoms and two nitrogen atoms. (c) $T_2^B T_2^B$ − termination (top edge) and $T_2^N T_2^N$ − termination (bottom edge), comprising 200 nitrogen and boron atoms while the top right sub-figure shows the*

*relaxed boron-nitride NT sculpturene with top line contains four-membered boron rings (four), heptagons (three), horizontal pentagons pairs (four), pentagons (three) and horizontal hexagons pairs (four), the bottom line contains hexagon rings of nitrogen (four), ten-membered boron and nitrogen ring (three) and horizontal pentagons pairs (eight). For clarity, in the defect lines all right sub- figures (a-c), boron is shown in green and nitrogen is shown in red.*

In contrast with the wide band gap of BN, the BN sculpturenes shown in figure 10 possess hugely-reduced gaps [18] of approximately 1.25 eV (figure 11a-c), which is similar to that of silicon (1.12 eV) [19]. Furthermore the BNNT sculpturenes in figures 11b and 11c possess additional energy gaps in the energy window from -2 eV to 2 eV.

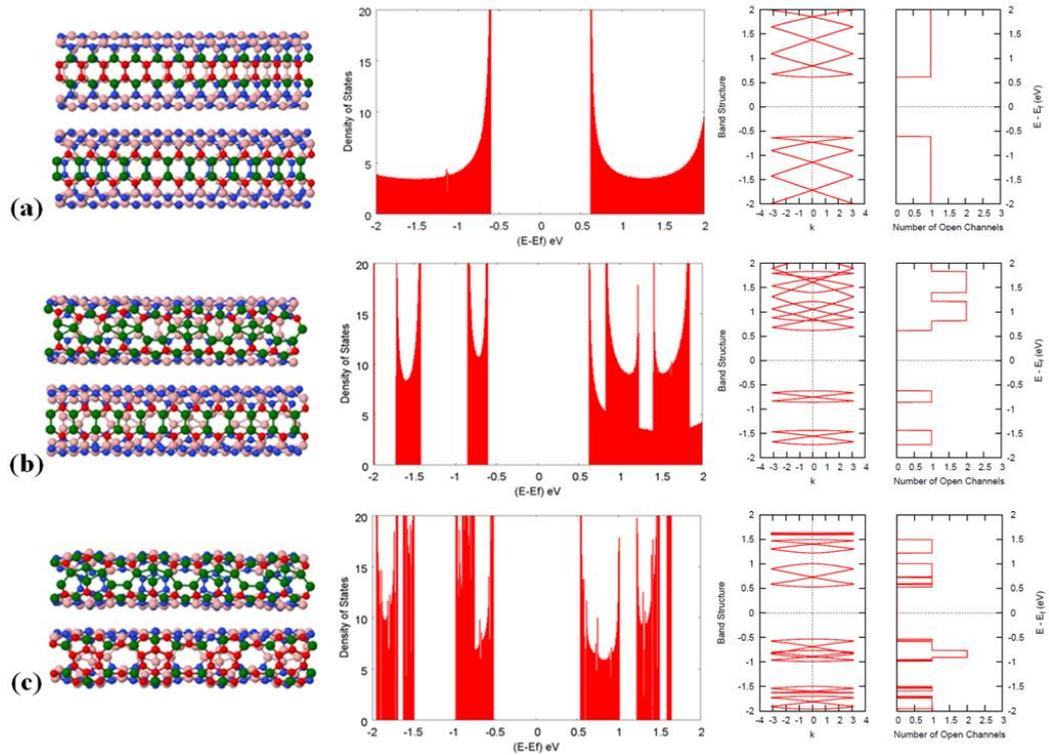

*Figure 11: The left sub-figures (a-c) Show BNNT sculpturenes (a) A BNNT sculpturene with two zips with along the longitudinal direction with two lines of hexagonal ring, the top line contains hexagon rings zip, each hexagonal ring contains four nitrogen atoms (red) and two boron atoms (green) while the bottom line contains hexagon rings zip, each hexagonal ring contains four boron atoms (green) and two*

*nitrogen atoms (red). (b) A BNNT sculpturene, the top line contains four-membered boron (three) rings, heptagons (four), horizontal pentagons pairs (three), pentagons (four) and horizontal hexagons pairs (three)while the bottom line contains hexagon rings zip with four boron atoms and two nitrogen atoms.(c) A BNNT sculpturene with a top line contains four-membered boron rings (four), heptagons (three), horizontal pentagons pairs (four), pentagons (three) and horizontal hexagons pairs (four) and the bottom line contains hexagon rings of nitrogen (four), ten-membered boron and nitrogen ring (three) and horizontal pentagons pairs (eight). The middle and right sub-figures (a-c) show the density of states, band structures and number of open channels respectively. The calculations carried out with 3000 k-points.*

The number of open channels for each system is shown in the far right sub-figures in figure 11. These show that within the energy window from -2 eV to 2 eV, there is only one open channel for the sculpturenes in figure 11a and one or two open channels for the others. Since the band gap is decreased in the zipped nanotubes, these can support current at low-bias voltages, which would not be possible in the insulating (6,6) BNNT. The I-V plots (figure 12) show that the current is enhanced compared to an ideal BNNT.

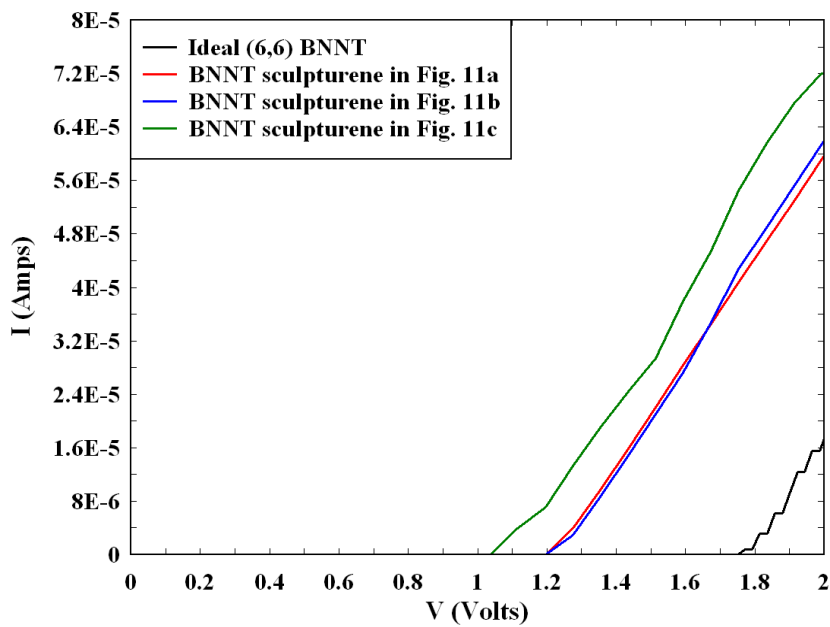

*Figure 12 : Shows I-V curve of the BNNT sculpturenes in figures 11 (a, b, and c).*

## 4. Electronic properties of carbon-BN hetero-sculpturenes formed from straight nanoribbons.

In this section, we examine the electronic properties of heteronanotubes obtained by relaxing bilayer ribbons formed from a monolayer of carbon on top of a monolayer of BN. As an example, figure 13 shows the hetero-nanotube obtained by sculpting a hetero-bilayer from a monolayer of graphene on top of a single layer of boron nitride and allowing reconstruction. The resulting nanotube consists of a half cylinder of carbon joined to a half cylinder of boron nitride.

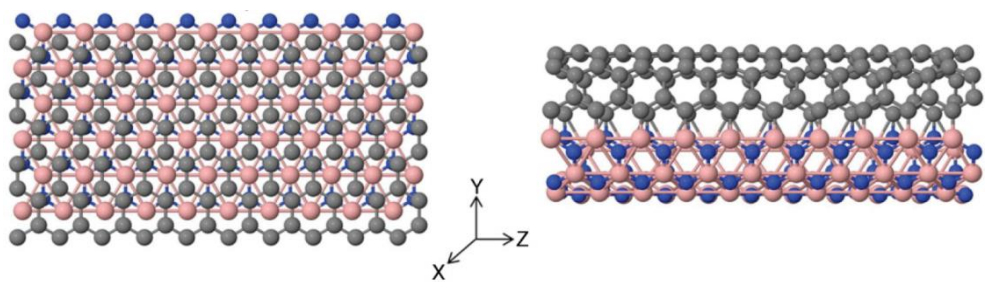

*Figure 13 Left; a Super cell of hetero-bilayer nanoribbons of boron nitride/graphene (periodic in z direction) which contains 240 atoms while the right sub-figure shows the relaxed (6,6) armchair hetero-NT obtained by relaxing the structure on the right.*

To understand the electronic structure of such hetero-sculpturenes, it is useful to examine their properties as a function of the relative widths of the carbon and BN sections of the NT. Such structures are likely to be challenging to realise experimentally, although in-plane junctions between monolayer BN and graphene are known [20] and therefore one can envisage cutting hetero-bilayer ribbons whose upper layer is formed in part from graphene and in part from BN, as shown in figure 14. This figure shows a range of resulting hetero-nanotubes, starting from (a) a perfect (6,6) BNNT, then containing progressively thicker sections of carbon (b-f) and ending with a perfect (6,6) CNT. In all cases, the hetero nanotubes have a boron-carbon and a nitrogen-carbon interface and the overall structure is made of hexagonal rings.

These structures are useful for illustrating the role of the interface between graphene and BN, because they share features associated with impurities in carbon nanotubes [21-26], specifically those associated with the doping of CNTs with boron [16, 26] and nitrogen [13, 23]. For such impurities, a characteristic peak near the Fermi energy is found, with the peak associated with boron below the Fermi energy and the peak associated with nitrogen above the Fermi energy. Similar peaks appear in the DOS presented in figure 14 (c, d, e and f). Beginning with the well-known DOS of the perfect (6,6) CNT (figure 14g), we see that the addition of a one-ring-thick nanoribbon of boron-nitride to the structure (figure 14f) creates a new feature near the Fermi energy. As the thickness of the boron-nitride strip increases (figures 14 e to c), the feature persists. Since the feature has two peaks (one above and one below the Fermi energy), it is reasonable to expect the peaks are associated with either the boron-carbon interface or the nitrogen-carbon interface. To demonstrate this, figure 15 shows the local density of states (LDOS) for the structure of figure 14d, centred on each peak. The orbital distributions shown in figure 15 were obtained by calculating the LDOS in a 0.04 eV of energy window centred on the two peaks shown in figure 14d.  Clearly, the peaks are localised at the two interfaces. Figure 14b shows that adding a one-ring-thick graphene nanoribbon to a BN nanotube significantly reduces the gap to approximately 0.8 eV, while the addition of a two-ring-thick graphene nanoribbon closes the gap completely. Interestingly, the two-peak feature associated with the boron-carbon and nitrogen-carbon interfaces is robust, being present even in figure 14b.

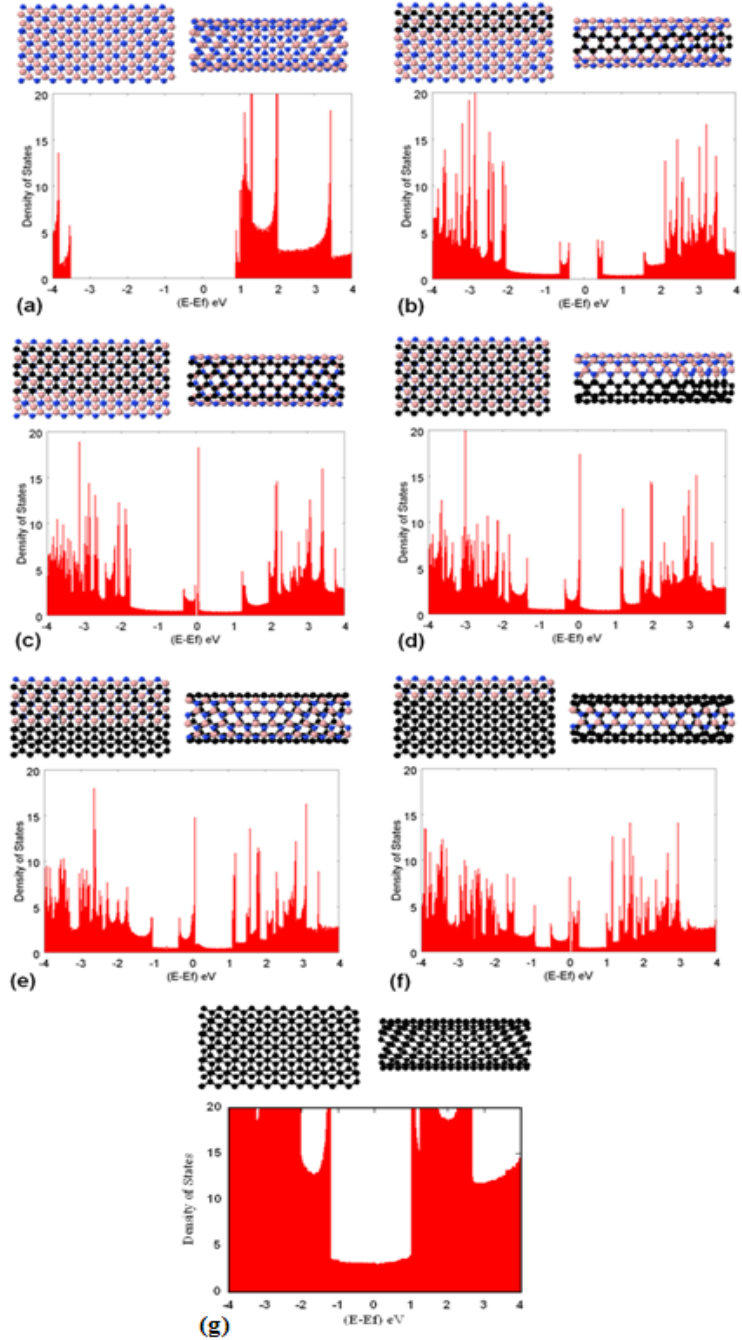

*Figure 14: Figures (a) and (g) the top left sub-figures show AB-stacked bilayer boron nitride and AB-stacked graphene nanoribbons which are relaxed to ideal BN and carbon nanotube sculpturenes (the top right sub-figures) respectively. (b) The top left sub-figures show AB-stacked hetero bilayer nanoribbon (the top layer with one-ring-thick graphene stacked on top of boron nitride nanoribbon), the structure relaxed to hetero nanotube with one-ring-thick graphene (the top right sub-figure). (c) The top*

*left sub-figures show AB-stacked hetero bilayer nanoribbon (the top layer with two-ring-thick graphene stacked on top of boron nitride nanoribbon), the structure relaxed to hetero nanotube with two-ring-thick graphene (the top right sub-figure). (d) The top left sub-figures show AB-stacked hetero bilayer nanoribbon (the top layer is graphene nanoribbon stacked on top of boron nitride nanoribbon), the structure relaxed to hetero nanotube consist of a half cylinder of carbon joined to a half cylinder of boron nitride (the top right sub-figure). (e) The top left sub-figures show AB-stacked hetero bilayer nanoribbon (the top layer graphene nanoribbon stacked on top of boron nitride nanoribbon with one-ring-thick graphene), the structure relaxed to hetero nanotube with four-ring-thick graphene (the top right sub-figure). (f) The top left sub-figures show AB-stacked hetero bilayer nanoribbon (the top layer graphene nanoribbon stacked on top of boron nitride nanoribbon with two-ring-thick graphene), the structure relaxed to hetero nanotube with five-ring-thick graphene (the top right sub-figure). All sub-figures on the bottom show the DOS of the hetero nanotube sculpturenes carried out with 3000 k-points.*

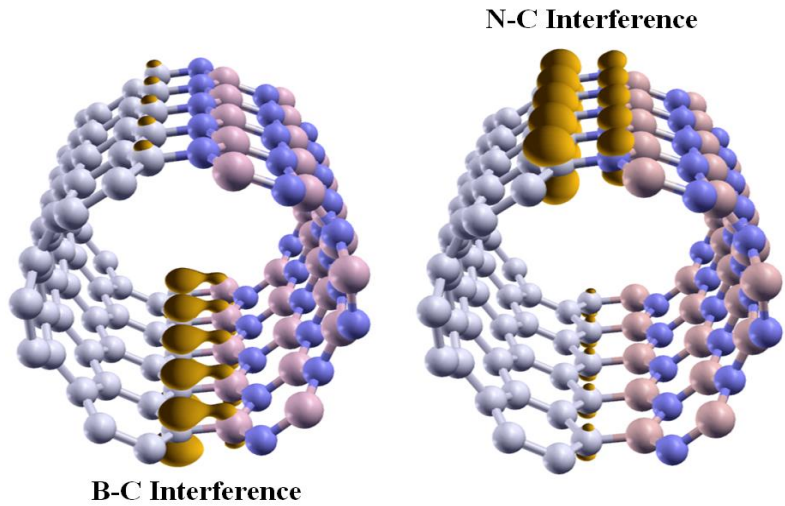

*Figure 15: The left sub-figure shows that the local density of states (in yellow) of the scultpturene of figure 14d at energy E = ???? is concentrated at the Boron-Carbon interface (in yellow), while the local density of states at E= ??? shown in the right sub-figure is concentrated at the of Nitrogen-Carbon interface.*

The band structures and number of open channels for these hetero nanotubes are shown in figure 16. With the obvious exceptions of the perfect BNNT and CNT and the one-ring-thick graphene nanoribbon structure which possesses a band gap, the electronic structures are remarkably similar. As a consequence the I-V curves in figure 17 are also similar for all of these hetero nanotubes, with the exception of the sculpturenes of figure 14b.

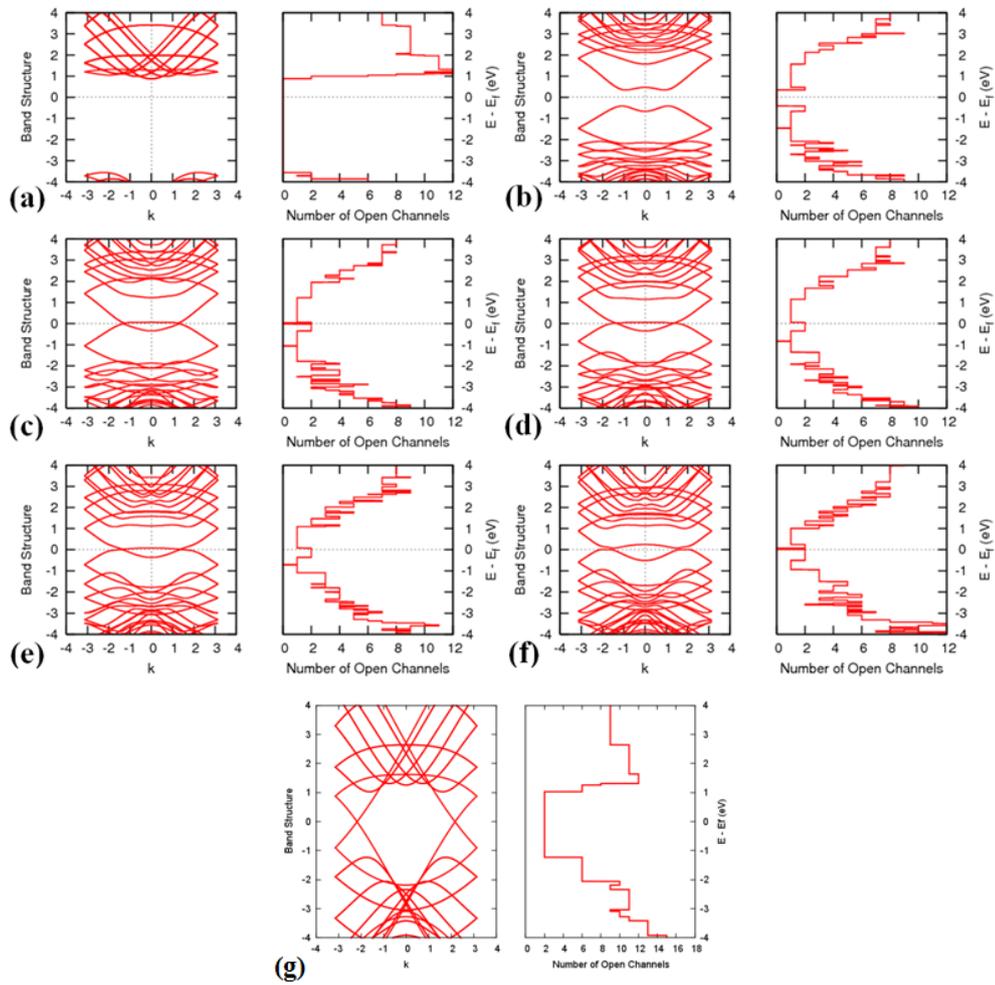

*Figure 16: figures (a-g) Show the band structure and number of open channels for the hetero nanotube sculpturenes shown in figures 14 (a-g) respectively. The calculations carried out with 3000 k-points.*

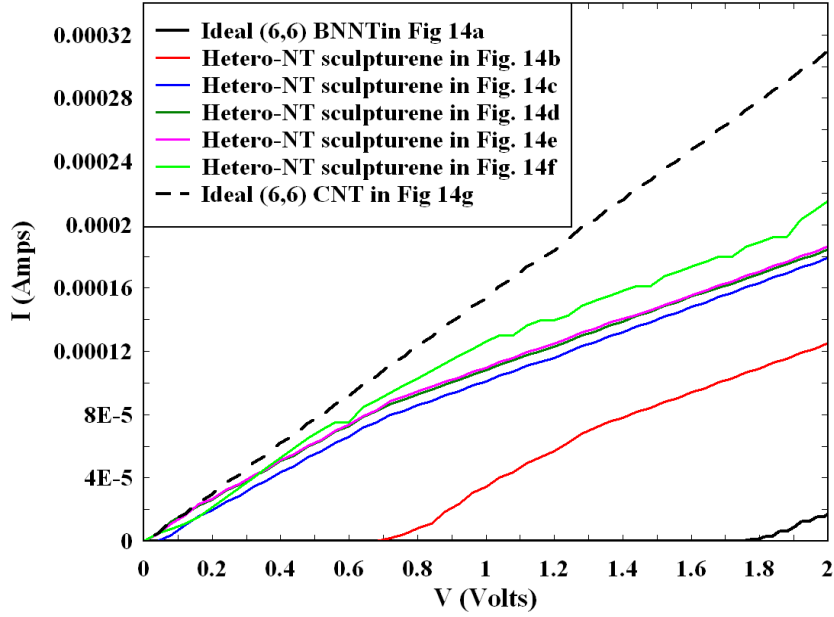

*Figure 17: Shows I-V curve of the BNNT sculpturenes in figures 14 (b, c, d and f).*

## 5. Electronic properties of in situ sculpturenes.

One of the biggest challenges for future nanoelectronics is the design of reliable methods to connect sub-10 nm devices to electrodes. As illustrated in figures 18 and 23, sculpturenes have the potential to overcome this contacting problem. As an example, figure 18 shows that by sculpting a narrow rectangular BiGNR between two wider BiGNRs and allowing them to relax, the resulting sculpturene is a CNT automatically connected to bilayer graphene electrodes. As a second example, figure 23 shows that more complex structures, such as a CNT torus connected to electrodes can also be obtained. The structures of figures 18 and 23 comprise a central scattering region attached to periodic electrodes on the left and right and therefore their transmission coefficients T(E) can be obtained using the non-equilibrium Green's function code SMEAGOL [27, 28], which utilises the DFT-based hamiltonian from SIESTA. In what follows, we investigate the transport properties of such sculpturene-based systems, which are constructed by sculpting a finite AB-stacked BiGNR (with the same terminations shown in figures 4a, 4b, 4c, 4d, 4e and 4f ) whose ends are connected to periodic bilayer graphene electrodes.

Figure 18a shows the transmission probability of sculpturene obtained by sculpting a nanoribbon from AB-stacked BiG with $T_1T_1/T_1T_1$ termination, while figures 18b and 18c show the transmission probability of zipped sculpturenes obtain by sculpting nanoribbons from AB-stacked bilayer graphene with $T_1T_1/T_2T_2$ and $T_1T_1/T_1T_2$ terminations respectively and then relaxing the initial structures to form CNT sculpturenes which are automatically connected to AB-stacked BiG electrodes.

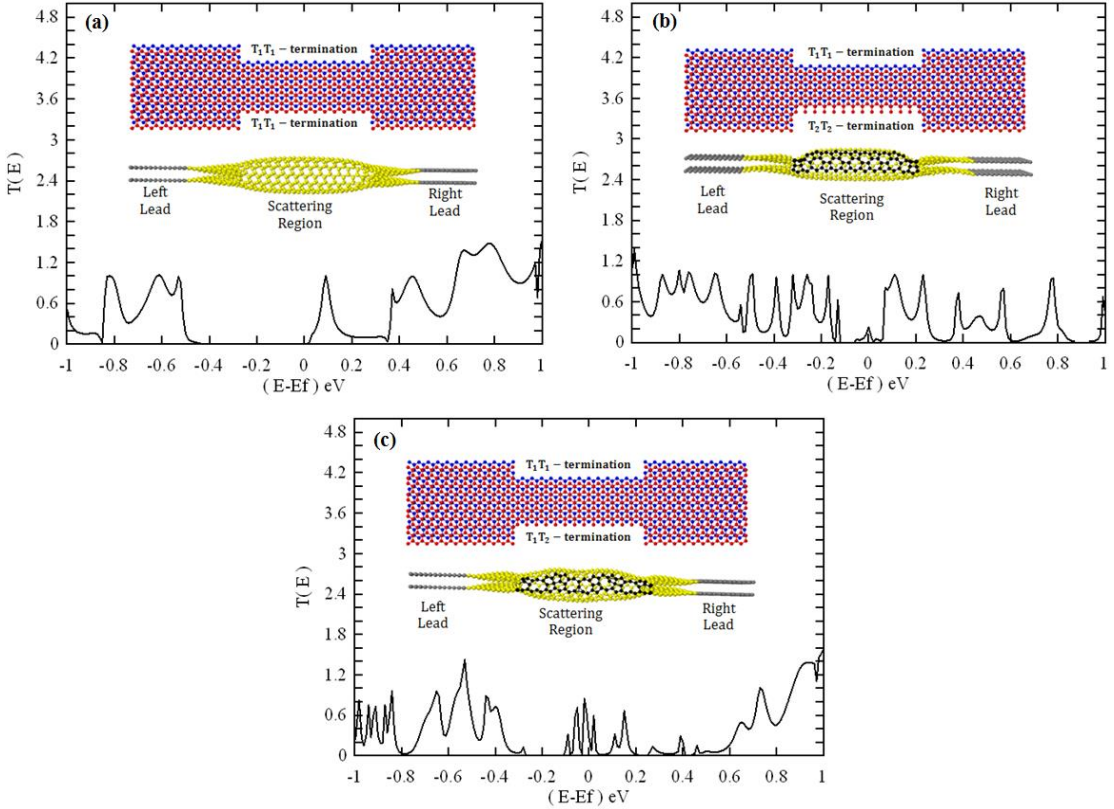

*Figure 18: the inset figures show, (a) The initial supercell (top) which contains a sculpted nanoribbon region with $T_1T_1/T_1T_1$-termination whose ends connected to AB-stacked BIG, the resulting sculpturene is shown underneath the initial supercell (1224 carbon atoms). (b) The initial supercell (top) which contains a sculpted nanoribbon region with $T_1T_1/T_2T_2$-termination whose ends connected to AB-stacked BIG, the resulting sculpturene is shown underneath the initial supercell (1198 carbon atoms). (c) The initial supercell (top) which contains a sculpted nanoribbon region with $T_1T_1/T_1T_2$-termination whose ends connected to AB-stacked BIG, the resulting sculpturene is shown underneath the initial supercell (1211 carbon atoms). The black*

*curves represent the zero bais transmission probability of the CNT sculpturenes which are connected to BiG electrodes after relaxation.*

Figure 18a shows that compared with the T(E) of the (6,6) CNT obtained from the $T_1T_1/T_1T_1$ terminations shown in figure 5a, there is a significant reduction in T(E) at the Fermi energy, whereas there is an enhancement in the T(E) at Fermi energy of the sculpturenes in figures 18b and 18c, which contain lines of non-hexagonal rings and correspond to the $T_1T_1/T_2T_2$ and $T_1T_1/T_1T_2$ terminations of fig 7a and figure 7d.

To illustrate the origin of these features in T(E), the local densities of states (LDOS) of the sculpturenes in figures 18a and 18b are shown in figures 19 and 20 respectively, within narrow energy ranges of interest near the Fermi energy.

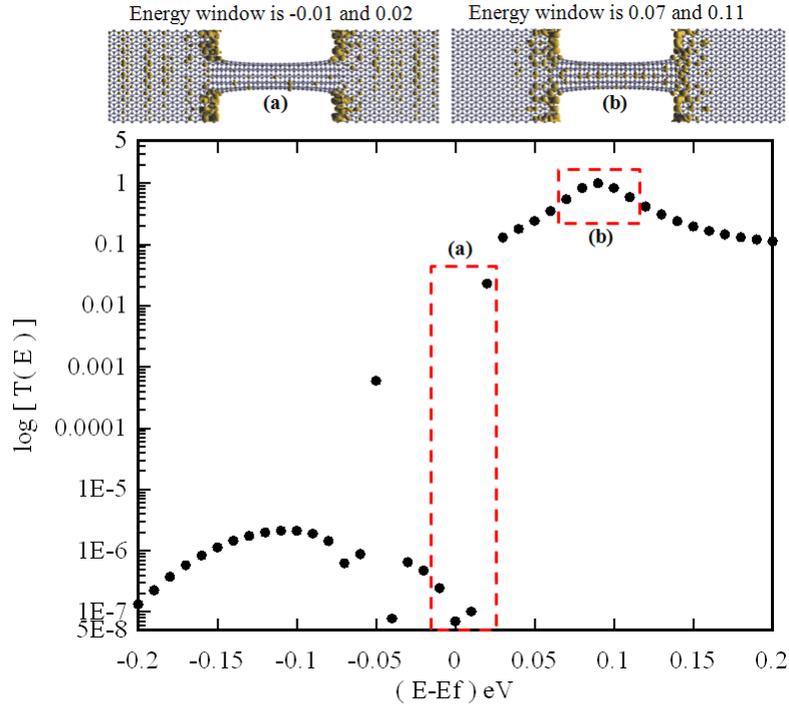

*Figure 19: Figures (a) and (b) show the LDOS of the sculpturene shown in figure 18a. The bottom figure shows the transmission probability in logarithm scale; the red dotted rectangles highlighted the energy ranges which are corresponding to the energy ranges of the LDOS.*

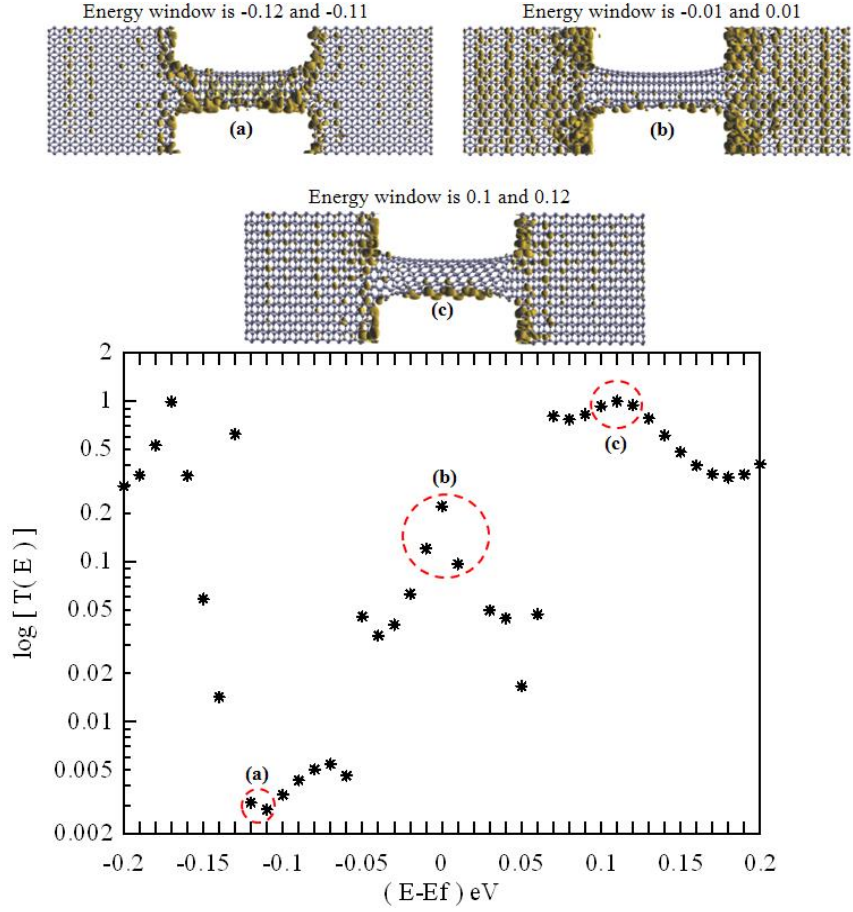

*Figure 20: figures (a), (b) and (c) show the LDOS of the sculpturene shown in figure 18b. The bottom figure shows the transmission probability in logarithm scale; the red dotted circles highlighted the energy ranges which are corresponding to the energy ranges of the LDOS.*

Figure 19 shows that the low value of T (E) near the Fermi energy for the sculpturene of figure 18a is associated with the absence of states in the central CNT region at this energy, whereas the higher transmission in energy region *b* arises from the presence of an extended state within the central CNT. In contrast, figure 20 shows that the LDOS at and around Fermi energy of the CNT sculpturene of figure 18b is non-zero along the lines of 5/7 rings, which leads to the relatively-high conductance of this sculpturene system.

Figures 21a, 5,16b and 21c show the transmission probability of quasi-disordered sculpturenes (with two lines of non-hexagon rings) which are obtained by sculpting a nanoribbon from AB-stacked BiG with $T_2T_2/T_2T_2$, $T_1T_2/T_1T_2$ and $T_2T_2/T_1T_2$ terminations respectively and allow them to reconstruct.

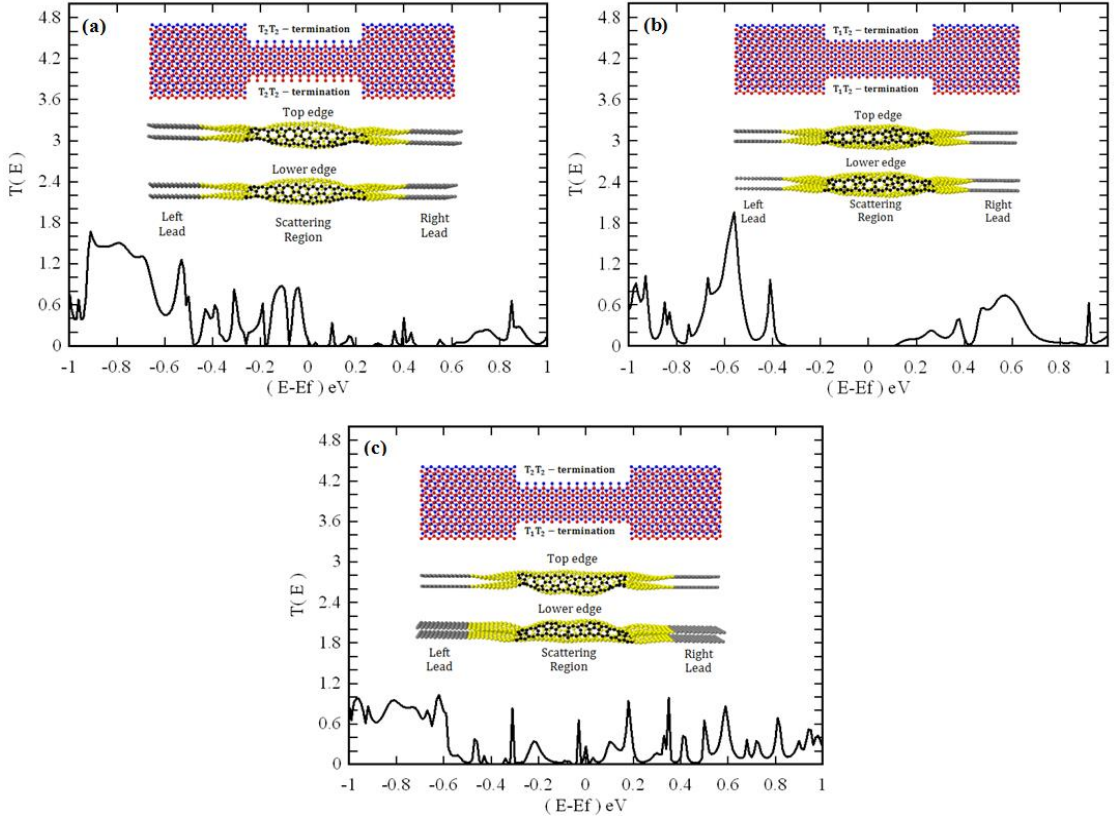

*Figure 21: the inset figures show, (a) The initial supercell (top) which contains a sculpted nanoribbon region with $T_2T_2/T_2T_2$-termination whose ends connected to AB-stacked BIG, the resulting sculpturene is shown underneath the initial supercell (1172 carbon atoms). (b) The initial supercell (top) which contains a sculpted nanoribbon region with $T_1T_2/T_1T_2$-termination whose ends connected to AB-stacked BIG, the resulting sculpturene is shown underneath the initial supercell (1198 carbon atoms). (c) The initial supercell (top) which contains a sculpted nanoribbon region with $T_2T_2/T_1T_2$-termination whose ends connected to AB-stacked BIG, the resulting sculpturene is shown underneath the initial supercell (1185 carbon atoms). The black curves represent the zero bais transmission probability of the CNT sculpturenes which are connected to BiG electrodes after relaxation.*

Like the structures of figures 18b and 20, the non-hexagonal carbon rings which appear in the final sculpturenes again strongly influence transport. The left sub-figure of figure 22 shows that at low voltages, the sculpturenes shown in figure 18a and 18c can carry the higher currents, while at higher voltages between approximately 0.2V and 1.4V, the structure shown in figure 18b carries the higher current. Similarly the right sub-figure shows the sculpturene system with two lines of 5/7 rings (figure 21a) carries the higher current up to approximately 1V, whereas the lower current corresponds to the system shown in figure 21b.

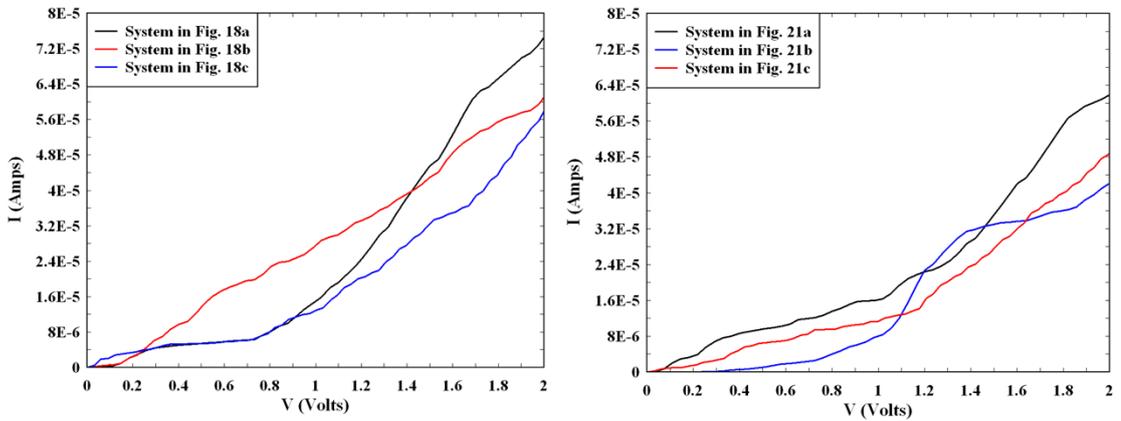

*Figure 22: Shows I-V curve of the Sculpturene systems are shown in figures 18 and 21. The left sub-figure shows the I-V curve of the systems in figures 18a, 18b and 18c. The right sub-figure shows the I-V curve of the systems in figures 21a, 21b and 21c.*

Over a range of energies near the Fermi energy, the value of T(E) shown in figures 18 and 21 never exceeds unity. To illustrate that this feature is shared by other in situ sculpturenes, we now compute T (E) for the torus of figure 23a and the carbon nanobud-like structure of figure 24a.

For the structure figure 23a, there are two open scattering channels in the energy range of from -0.8 eV to 0.8 eV and therefore in this range, the transmission matrix $\vec{t}$ is a 2x2 matrix of the form:

$$\vec{t} = \begin{pmatrix} t_{11} & t_{12} \\ t_{21} & t_{22} \end{pmatrix} \qquad (2)$$

The transmission coefficient T (E) is therefore equal to the sum of the two eigenvalues of the transport matrix τ, given by:

$$\tau = \vec{t}\vec{t}^{\dagger} \qquad (3)$$

Figure 23b shows a plot of T(E), while figure 23c shows plots of the two eigenvalues. One of the two eigenvalues is dominant (red curve) whereas the second eigenvalue (black curve) is negligible.

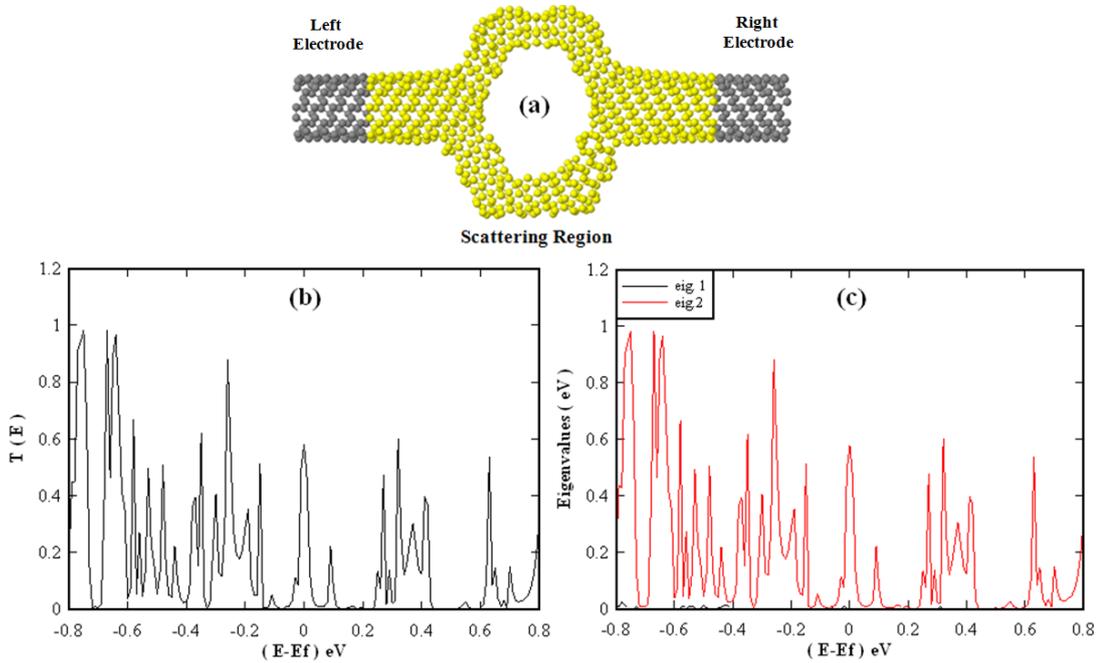

*Figure 23: (a) Shows torus sculpturene system. (b) Shows the zero bias transmission probability of torus sculpturene. (c) Shows the two eigenvalues which are correspond to the two open channels in the right lead, the red curve represents the dominant eigenvalue and the black curve is the non-dominant eigenvalue.*

The presence of a single dominant transmission eigenvalue is also found in of the carbon nanobud-like structure shown in figure 24a, whose T (E) is shown in figure 26b. Figure 24c again shows that this sculpturene possesses single dominant eigenvalue (red curve) and a negligible second eigenvalue (black curve).

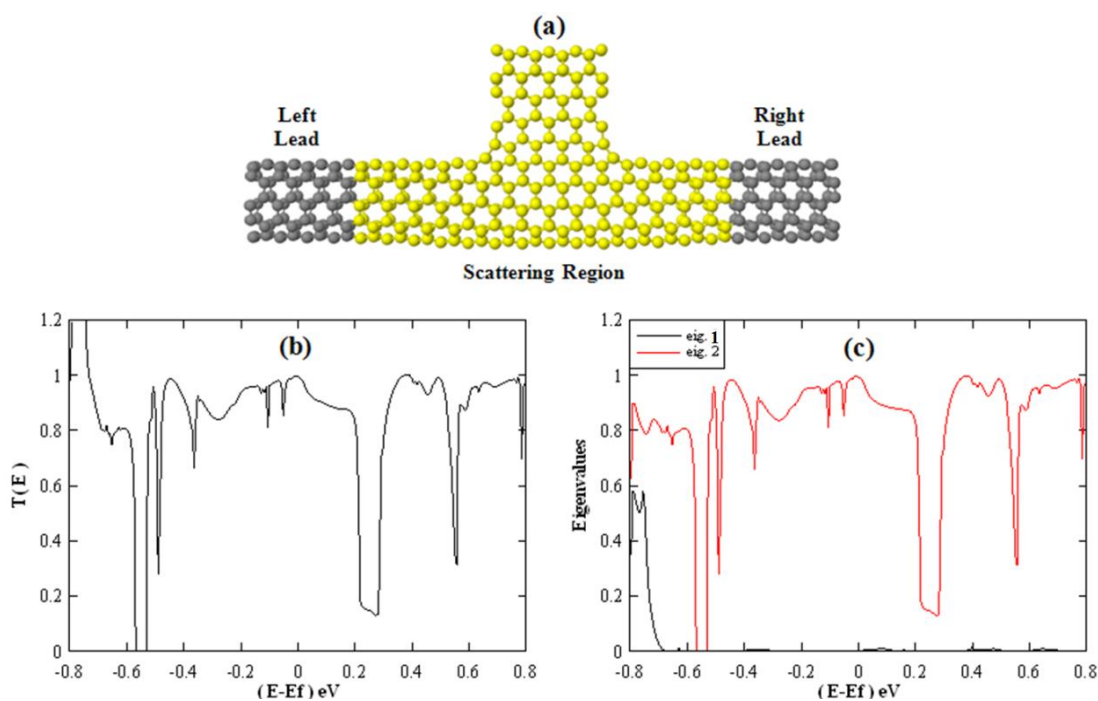

*Figure 26: (a) Shows the nanobud sculpturene. (b) Shows the zero bias transmission probability of the nanobud sculpturene. (c) Shows the two eigenvalues which are corresponding to the two open channels in the leads, the black curve shows the non-dominant eigenvalue and the red curve shows the dominant eigenvalue.*

The presence of a single dominant channel, with T(E) never exceeding unity is typical of electron transport through single molecules, where T(E) exhibit Breit-Wigner and/or Fano resonances when E coincides with molecular energy levels (shifted by the self-energy of the electrodes)[42] . Hence the sculpturenes of figures 18-26 share the advantages of single-molecule devices associated with the presence of transport resonances, but do not suffer from the well-known disadvantages associated with highly-resistive contacts to the electrodes. Like single-molecule devices [43], transport through the above structures is also controlled by geometry, but on a significantly-higher length scale.

## 6. Summary

The electronic properties of all-carbon, BN and hetero sculpturenes have investigated.

For the simplest sculpturenes formed from reconstructed BiGNRs, the resulting nanotubes depend not only on the chirality of the initial NRs, but also on the combination of edge terminations. Typically the sculpturene nanotubes possess a unit cell which is longer than the initial BiGNR and in many cases possess lines of non-hexagonal rings, which as shown in figure 7, lead to a reduction or even elimination of energy gaps in the DOS of such structures. For BN sculpturenes, figure 11 shows that the energy gap near the Fermi energy is found to persist, but is significantly reduced for the bulk value. In the case of hetero-nanotube sculpturenes formed from reconstructed ribbon containing both carbon and BN strips, pronounced peaks in the DOS are found, which arise from carbon-boron and carbon-nitrogen bonds. In situ sculpturenes connected to electrodes have also been investigated, including nanotube-like structures, a torus and a nanobud. In all cases, electron transport was found to be dominated by a single scattering channel, leading to electrical conductance less than or of order the conductance quantum.

In addition to the sculpturenes considered here, one can envisage many other examples, including sculpturenes formed from other 2d materials and combinations thereof. In addition, under realistic conditions, the initially-cut bilayer structures may have disordered edges, which would add to the variety of achievable sculpturenes. On the other hand, a molecular dynamics study of defective carbon nanotubes at 1500K has shown an interesting self-healing property [29], which suggests that, at moderate temperature, the disorder in these nanotubes may "heal" into a smaller set of stable structures of the type considered in this paper. For the future it would be of interest to the self-healing properties of sculpturenes and their atomic-scale dynamics during simultaneous cutting and reconstruction.